\documentstyle[12pt,epsf,epsfig,psfig]{article}
\def\J{$J/\psi$}
\def\j{J/\psi}
\def\X{$\chi_c$}
\def\x{\chi}
\def\P{$\psi'$}

\def\U{$\Upsilon$}

\def\C{c{\bar c}}

\def\e{\epsilon}

\def\S{\sigma_{\C}}

\def\be{\begin{equation}}
\def\ee{\end{equation}}

\def\lsim{\raise0.3ex\hbox{$<$\kern-0.75em\raise-1.1ex\hbox{$\sim$}}}
\def\gsim{\raise0.3ex\hbox{$>$\kern-0.75em\raise-1.1ex\hbox{$\sim$}}}

\def\NP{{ Nucl.\ Phys.\ }}
\def\PL{{ Phys.\ Lett.\ }}
\def\PR{{ Phys.\ Rev.\ }}

\def\PRL{{ Phys.\ Rev.\ Lett.\ }}

\def\ZP{{ Z.\ Phys.\ }}

\begin{document}

\thispagestyle{empty}

\noindent September 9, 2000~ \hfill BI-TP 2000/31

\vskip 2.5 cm

\centerline{\Large{\bf The Search for the QGP:}}

\bigskip

\centerline{\Large{\bf A Critical Appraisal}\footnote{Invited talk at
{\sl Lattice 2000: XVIII International Symposium on Lattice Field
Theory}, August 17-22, 2000, Bangalore/India; Proceedings to appear 
in\NP B. Also presented at the {\sl RHIC and AGS
Users Meeting}, August 7-8, 2000, Brookhaven National Laboratory,
Upton/NY, USA, and at {\sl Bologna 2000: Structure of the Nucleus at the
Dawn of the Century}, May 29-June 3, 2000, Bologna/Italy.}}

\vskip 1.0cm

\centerline{\bf Helmut Satz}

\bigskip

\centerline{Fakult\"at f\"ur Physik, Universit\"at Bielefeld}
\par
\centerline{D-33501 Bielefeld, Germany}

\vskip 1.5cm

\noindent

\centerline{\bf Abstract:}

\medskip

Over the past 15 years, an extensive program of high energy nuclear
collisions at BNL and CERN was devoted to the experimental search
for the quark-gluon plasma predicted by QCD. The start of RHIC
this year will increase the highest available collision energy by a
factor 10. This seems a good time for a critical assessment: what have
we learned so far and what can we hope to learn in the coming years?

\vskip 1.5cm












\newpage

{\parindent=0pt
{\it ``What's the good of Mercator's North Poles and Equators,
\par
Tropics, Zones and Meridian Lines?"
\par
So the Bellman would cry: and the crew would reply
\par
``They are merely conventional signs!"}

\medskip

Lewis Carroll,
\par
{\it The Hunting of the Snark}}

\vskip 1.5cm

\noindent{\bf 1.\ Expecting the Unexpected}

\bigskip

QCD predicts that at high energy density, hadronic matter will turn
into a plasma of deconfined quarks and gluons. Eighteen years ago,
theorists and experimentalists met in Bielefeld to chart the course for
the study of quark-gluon plasma formation in heavy ion collisions. It
was noted then: ``In the past decades, we have investigated the strong
interaction of elementary particles at higher and higher energies. So
far, we have not been able to study in parallel strongly interacting
macroscopic systems at higher and higher [energy] density. The
thermodynamics of very dense matter is largely an unexplored field
for terrestial experimental physics" \cite{Jacob}.

\par

In 1986/1987, experimental studies of high energy nuclear collisions
began at the BNL-AGS and the CERN-SPS. The AGS started with $^{28}$Si
beams (at 15 GeV/nucleon) and later went on to $^{200}$Au beams (at 12
GeV/nucleon) incident on heavy targets. At CERN, $^{16}$O and $^{32}$S
beams (at 200 GeV/nucleon) were followed in 1995 by $^{208}$Pb beams
(at 160 GeV/nucleon). While both labs have by now provided a wealth of
data on the hadronic final states produced in the different collisions,
the AGS program did not include any experiment able to reach back into
the early `hot' stages of the collision evolution. Moreover, estimates
in general indicate that with the AGS beam energy one can at best barely
reach the initial energy densities required for quark-gluon plasma
formation. For these reasons, most of our present knowledge about a
possible onset of deconfinement in nuclear collisions has come from the
CERN-SPS.

\par

At the beginning, the search for colour deconfinement in high energy
nuclear collisions was very much an exploratory endeavor. As in most
explorations, the enthusiasm of the explorers was greater than their
knowledge of what to expect or look for. There was a general feeling
that if the quark-gluon plasma was indeed produced, it would manifest
itself in a variety of unknown but dramatic ways, including the end of
the world; this, however, was ruled out \cite{Hut}. So the basic idea
in the planning of experiments was to cover all possibilities, hoping
that in this way the QGP would be observed somehow, somewhere,
preferably everywhere.

\par

Today, we are somewhat wiser. Changing conditions generally lead to a
change in the related observables; if this happens in a monotonic way,
it is indeed merely a conventional sign and has nothing to do with
any phase transition. What we are really looking for
is something like an `order parameter' in statistical mechanics:
an observable which remains constant over a range of different
conditions, but which then at a certain point starts changing. Such
non-analytic behaviour could be the consequence of a critical or
transition point. It is not yet known if an experimentally accessible
order parameter for deconfinement really exists.

\par

It is clear today, however, that hard and soft probes of nuclear
collisions provide information about different stages of the produced
state, determined by the scale of the probing phenomenon. For hard
probes, the associated length scale is much less than the confinement scale
$\Lambda_{\rm QCD}^{-1} \sim 1$ fm; hence these can be used to study
directly the short-distance features of a hot deconfined medium. Soft
probes are of hadronic size and appear when the density of the medium
has dropped sufficiently to allow the existence of hadrons. Hence they
naturally provide information about the hadronic state of the medium;
about earlier states they can at best give indirect information.

\par

The hadronisation of the QGP could lead to a hot interacting hadronic
medium, or it could directly result in free-streaming hadrons and/or
resonances. Hence there are three different stages to be considered:
the deconfined medium (QGP), a possible interacting hadronic medium,
and the final hadronic state.

\par

Let us recall the situation at the beginning of the experimental QGP
search some fifteen years ago. The main phenomena predicted to
constitute direct hard probes of the QGP were:
\begin{itemize}
\vspace*{-0.2cm}
\item{Hard thermal dilepton/photon emission \cite{Shuryak,K-M}; their
spectrum should serve as a thermometer of the thermal medium from which
they were emitted.}
\vspace*{-0.2cm}
\item{Jet quenching \cite{Bj-jet,Baier}; the energy loss of fast partons
passing through a deconfined medium should exceed that suffered in a
passage through hadronic matter.}
\vspace*{-0.2cm}
\item{\J~suppression \cite{Matsui,KS3}; colour screening in a deconfined
medium leads to quarkonium dissociation, while tightly bound heavy
quark states are not effected by a hadronic medium.}
\vspace*{-0.2cm}
\end{itemize}
\noindent
A possible interacting hadronic medium could be identified by
\begin{itemize}
\vspace*{-0.2cm}
\item{in-medium modifications of resonances observable in their dilepton
decays \cite{in-medium1,in-medium2}; in particular, changes of the
$\rho$ mass or width should be observable in this decay pattern.}
\vspace*{-0.2cm}
\end{itemize}
\noindent
The proposed soft probes of the hadronic final state were:
\begin{itemize}
\vspace*{-0.2cm}
\item{Strangeness enhancement \cite{RM-82,R-82}; a hot QGP contains
strange and the different non-strange quark species in almost equal
amounts, which, if preserved up to hadronization, should result in more
abundant strangeness production than observed in p-p interactions.}
\vspace*{-0.2cm}
\item{Transverse momentum broadening and flow \cite{VH,flow}; compared
to p-p interactions, a hot initial QGP could lead to more pronounced
expansion and/or specific expansion patterns.}
\vspace*{-0.2cm}
\end{itemize}
\noindent
In addition, some more or less exotic states (strangelets, disoriented
chiral condensates) or phenomena (softest point, tricritical point,
event-by-event fluctuations) were proposed at various times as QGP or
transition signatures; so far, none of these were observed.

\par

Thermal dileptons and photons were the subject of intensive experimental
studies. Nevertheless, there is up to now no real evidence for their
production. The observed intermediate mass dilepton enhancement
\cite{Kluberg} may turn out to be a first indication of thermal
dileptons \cite{Rapp}, but so far its origin remains completely open.
Moreover, real or virtual thermal photons, if observed, indicate the
temperature of the medium, not its confinement status.

\par

Much interesting theoretical work on jet quenching has appeared in
recent years (see \cite{Baier} for a survey); but energy limitations put
the effect out of reach for AGS and SPS. Hopefully it will hopefully
enter as a viable probe with the advent of RHIC and LHC.

\par

All other phenomena were indeed observed in high energy nuclear
collisions. However, we shall see that all occur in fact already under
conditions where presumably no QGP can be formed, e.\ g., in p-p or
p-A interactions. Hence their observation as such cannot be taken as
evidence for QGP production; it is crucial to specify if, when and
how the observed behaviour constitutes an unconventional sign. Various
studies over the past years have led to a systematic approach to this
problem, and we shall here try to apply it to the different observed
phenomena.

\par

Before doing that, however, let us briefly consider the thermodynamic
conditions necessary for QGP formation and compare them to what can be
achieved in nuclear collisions. From finite temperature lattice QCD we
know that deconfinement and chiral symmetry restoration occur in an
equilibrated medium of vanishing baryon number density at a temperature
of about 170 - 180 MeV; hence energy densities of some 1 - 3 GeV/fm$^3$
are required for QGP formation \cite{lattice}. Percolation theory
provides estimates valid also in non-equilibrium conditions; they lead
to similar values \cite{perco}.

\par

\begin{figure}[tbp]
\setlength{\unitlength}{1cm}
\begin{minipage}[t]{7.0cm}
\begin{picture}(6.5,6.5)
\hspace*{-0.2cm}
\psfig{file=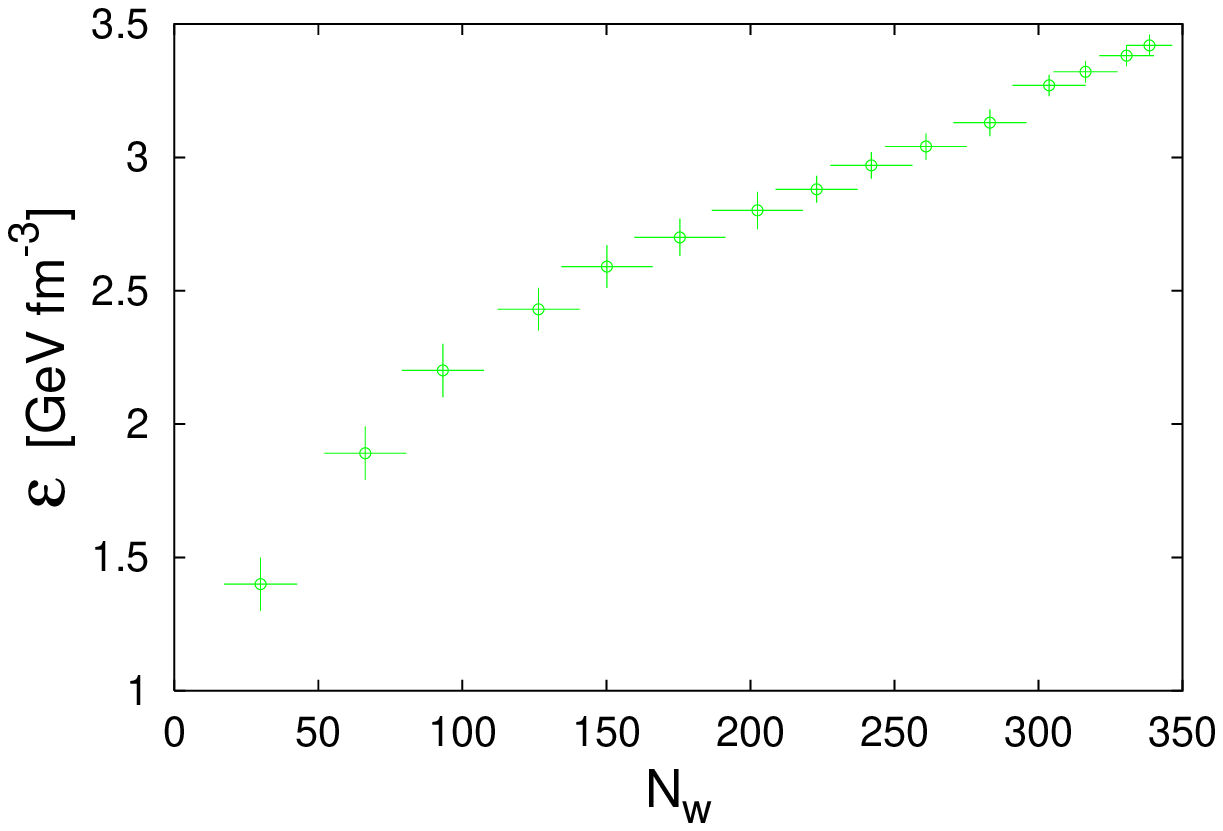,width=6.5cm,height=7cm}
\end{picture}\par
\caption{The variation of the initial energy density with the number
of participant nucleons \cite{NA50,Nardi}.}
\label{F1}
\end{minipage}\hfill
\begin{minipage}[t]{7.0cm}
\begin{picture}(7.5,7.5)
\hspace*{-0.2cm}
\psfig{file=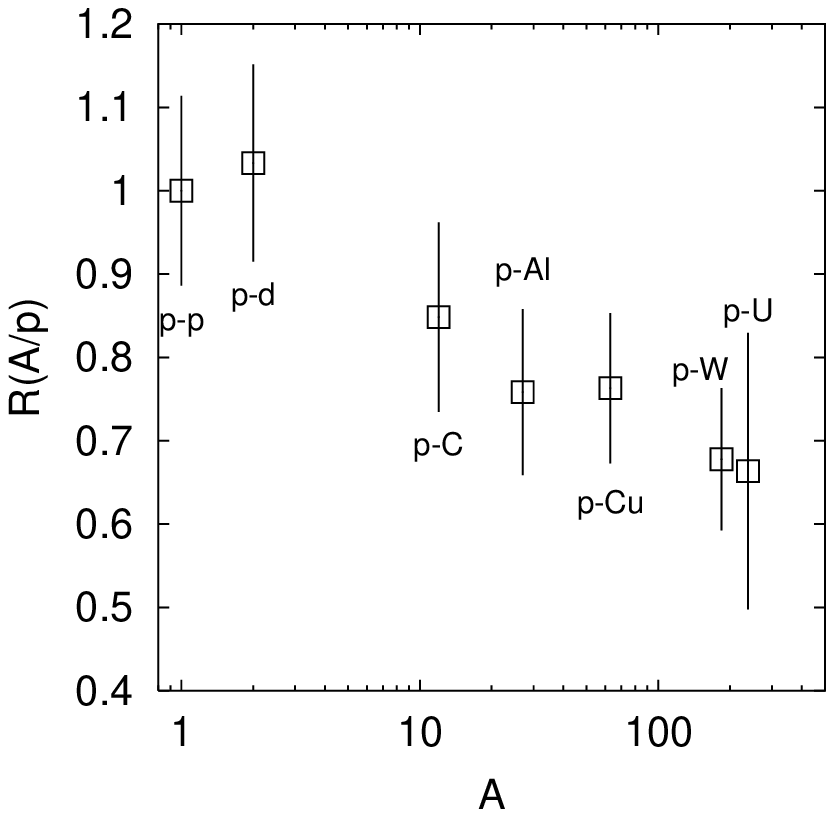,width=7.5cm,height=7cm}
\end{picture}\par
\caption{\J~suppression in p-A collisions \cite{NA38-f}.}
\label{F2}
\end{minipage}
\end{figure}

We want to compare the predicted critical conditions to those obtained
in nuclear collisions. The well-known evolution based on freely
streaming hadrons \cite{Bjorken} leads to the initial energy density
\be
\e_0 = {1 \over \tau_0 A_T} \left( {dE \over dy} \right)_{\!y=0},
\label{1.1}
\ee
where $A_T$ denotes the transverse overlap area of the two colliding
nuclei, $\tau_0 \simeq 1$ fm the formation time of the initial medium,
and $(dE/dy)_{y=0}$ the measured total energy of all the hadrons emitted
at central rapidity. Alternatively, $\e_0$ can be expressed in terms of
the measured central multiplicity $(dN/dy)_{y=0}$ and the average energy
$p_0$ of the hadrons at $y=0$, with $(dE/dy)_{y=0}=p_0(dN/dy)_{y=0}$.
The relevant quantities in Pb-Pb collisions were determined by several
SPS groups \cite{Baechler,NA50}. In Fig.\ \ref{F1}, we show the
resulting estimate of the initial energy density in SPS Pb-Pb
collisions as function of the number $N_w$ of participating nucleons.
Through a Glauber analysis, this can be related to the impact parameter
of the collision \cite{NA50,KLNS,Nardi}. We conclude that these collisions do
reach the regime in which deconfinement could be expected in the early
stages of the produced medium.

\bigskip

\noindent{\bf 2.\ Charmonium Suppression}

\bigskip

At present, the only accesible hard probe for a direct test of the
confinement status of the early medium is charmonium production. The
binding energy of the \J~is about $3~\!\Lambda_{\rm QCD}$ and its
radius about 0.2 fm, so that it indeed `sees' the short scales of a hot
pre-hadronic medium. In a deconfining environment, it dissolves once
the screening radius falls below its binding radius \cite{Matsui}. On a
microscopic level, short-distance QCD calculations show that \J-hadron
collisions at the hadron momenta attained in SPS collisions do not lead
to significant dissociation \cite{KS3}.
Gluons confined to hadrons are not sufficiently hard for a break-up;
deconfinement hardens the gluons and allows charmonium dissociation.
Hadronic models claiming a large \J~dissociation cross section in the
threshold region \cite{M-M} are not in accord with data from
\J~photoproduction \cite{Redlich}.

\par

The suppression of \J~production was observed quite early in the CERN
heavy ion program \cite{Baglin}; but it was soon noted that it occurs
as well already in p-A collisions (see Fig.\ \ref{F2}), where it
inceases with increasing A \cite{NA3}-\cite{NA38-f}. It thus depends
on the presence of a bulk medium, the nucleus; but the nucleus clearly
is a confined medium and hence \J~suppression cannot generally mean
deconfinement \cite{Gerschel}.

Before one can apply \J~suppression as probe in nuclear collisions, it
is therefore necessary to carry out a sequence of steps:
\begin{itemize}
\vspace*{-0.2cm}
\item{study the elementary production process in p-p collisions to
determine its behaviour in the absence of any bulk medium,}
\vspace*{-0.2cm}
\item{determine its behaviour in the confined nuclear medium given by
p-A collisions;}
\vspace*{-0.2cm}
\item{then check if in A-B collisions there are deviations from
the `normal' behaviour observed in the confined nuclear medium.}
\vspace*{-0.2cm}
\end{itemize}
\noindent
It should be emphasized that this procedure was by no means evident from
the beginning; it is a lesson in method learned in over ten years of
extensive experimental \J~production studies and the corresponding
theoretical attempts to understand the results (see \cite{Ramona,ROP}).
I want to suggest here that it in fact defines a more general approach
which any experimental QGP probe must follow in order to become
conclusive.

\par

\begin{figure}[hb]
\centerline{\epsfig{file=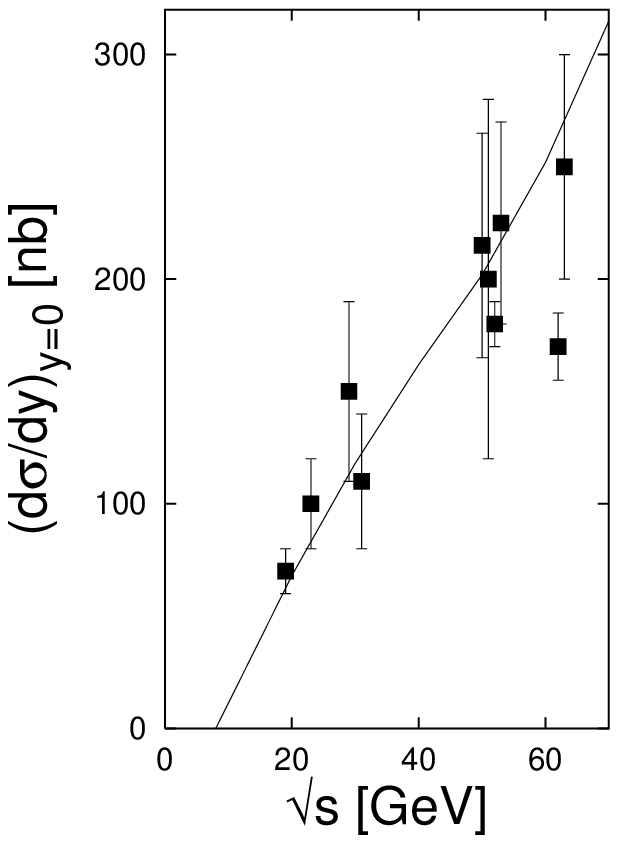,height=60mm,width=50mm}
\hspace*{2cm}
\epsfig{file=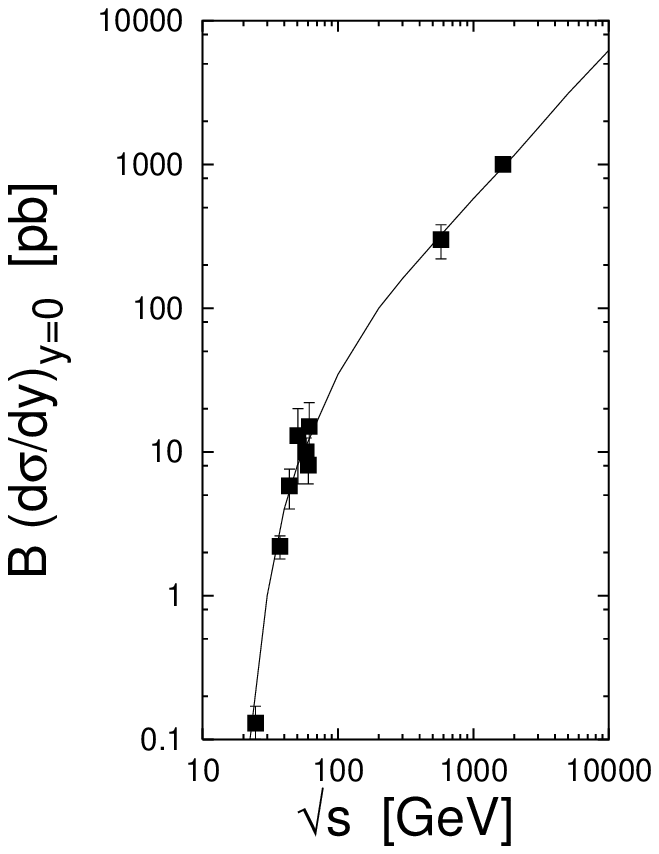,height=60mm,width=50mm}}
\caption{The energy dependence of \J~hadroproduction (left) and of
\U~hadroproduction (right) based on the colour evaporation model using
MRS D-' parton distributions \cite{Gavai}.}
\label{F3}
\end{figure}

\J~production in high energy nucleon-nucleon collisions starts with the
fusion of a target and a projectile gluon to form a colour octet $\C$
pair. In the colour field of the interaction region, the pair
neutralizes its colour to form a colour singlet state of \J~quantum
numbers. While the $\C$ production can be described perturbatively, the
colour neutralization is of non-perturbative nature. It can be described
conceptually by the colour evaporation model \cite{Gavai}, which is
based
on the total sub-threshold charm cross section. This is obtained by
integrating the perturbatively calculated $\C$ production over the mass
interval from $2m_c$ to $2m_D$,
\be
\S(s) = \int_{2m_c}^{2m_D} d\hat s \int dx_1 dx_2~g_p(x_1)~g_t(x_2)~
\sigma(\hat s)~\delta(\hat s-x_1x_2s), \label{2.1}
\ee
where $g_p(x)$ and $g_t(x)$ denote the gluon densities, $x_p$ and
$x_t$ the fractional momenta of projectile and target gluons, and
$\sigma$ the $gg \to \C$ cross section. The colour evaporation model now
states that the production cross section of any charmonium state $i$ is
given by
\be
\sigma_i(s)~=~f_i~\S(s), \label{2.2}
\ee
where $f_i$ is an energy-independent constant to be determined
empirically. This has two clear consequences: the energy dependence
of any charmonium production cross section is predicted to be that of
the perturbatively calculable sub-threshold charm cross section, and
the production ratios of different charmonium states must be
energy-independent. Both sets of predictions agree very well with data
over a considerable energy range. As illustration, we show
in Fig.\ \ref{F3} that the energy dependence is correctly described for
the \J, as it is in a corresponding treatment of \U~production, where
data are available up to 1.8 TeV. For further studies, see \cite{Gavai}.

\bigskip

While the colour evaporation model gives an empirically well supported
description of charmonium and bottonium production, it does not
provide a space-time description of the production process.
For charmonium production in p-A and A-B collisions, this
is crucial, however, and hence a more detailed model for colour
neutralization is needed. The colour octet model \cite{CO} proposes
that the colour octet $\C$ combines with a soft collinear gluon to form
a singlet $(\C\!-\!g)$ state. After a short relaxation time $\tau_8$,
this pre-resonant charmonium state (a `higher Fock space component' of
the \J) turns into the physical $\C$ singlet \J~mode (the `basic Fock
space component') by absorbing the accompanying gluon. Such a formation
process occurs as well for $\chi_c$ and \P~production. The colour octet
model encounters difficulties if the collinear gluons are treated
perturbatively, illustrating once more that colour neutralization seems
to require non-perturbative elements \cite{polarize}. However, it does
provide a conceptual basis for the evolution of the formation process.

\par

The life-time $\tau_8$ and the size $r_8$ of pre-resonant charmonium
can be estimated \cite{KS6}. Both are essentially determined by the
lowest momentum possible for confined gluons and hence are the same
for the different states \J, $\chi_c$ and \P, with $r_8 \simeq \tau_8
\simeq 0.20 - 0.25$ fm. This has striking consequences for
charmonium production in p-A collisions.

\par

\begin{figure}[tbp]
\setlength{\unitlength}{1cm}
\begin{minipage}[t]{7.0cm}
\begin{picture}(6.5,6.5)
\hspace*{-0.2cm}
\psfig{file=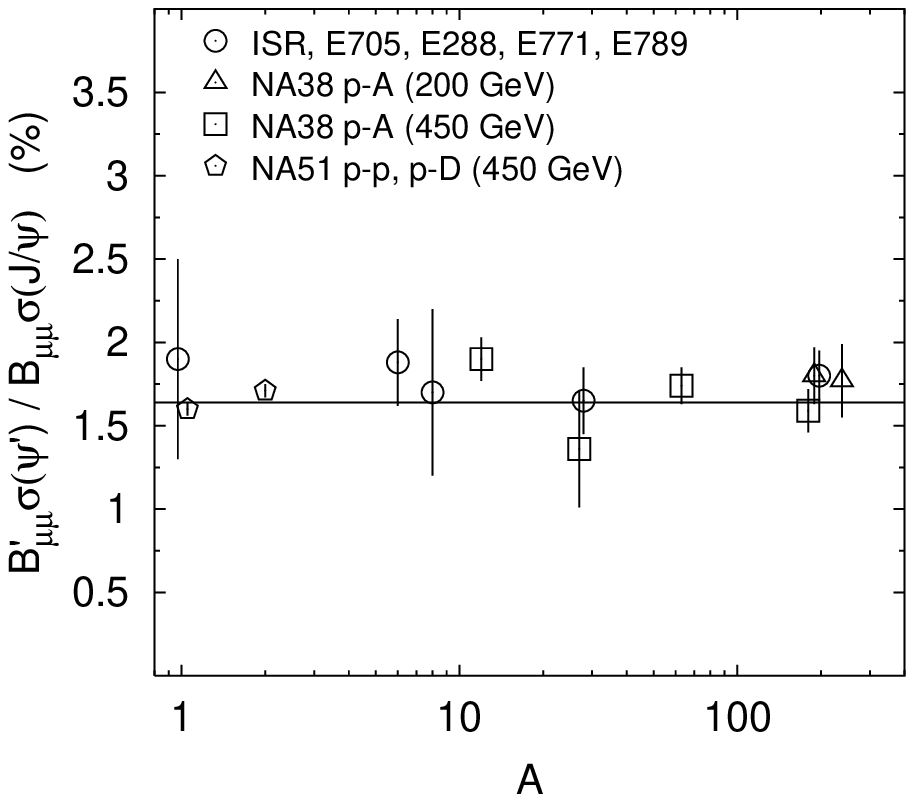,width=6.5cm,height=7cm}
\end{picture}\par
\caption{The relative A-dependence of \J~and \P~production in p-A
collisions \cite{Baglin}.}
\label{F4}
\end{minipage}\hfill
\begin{minipage}[t]{7.0cm}
\begin{picture}(7.5,7.5)
\hspace*{-0.2cm}
\psfig{file=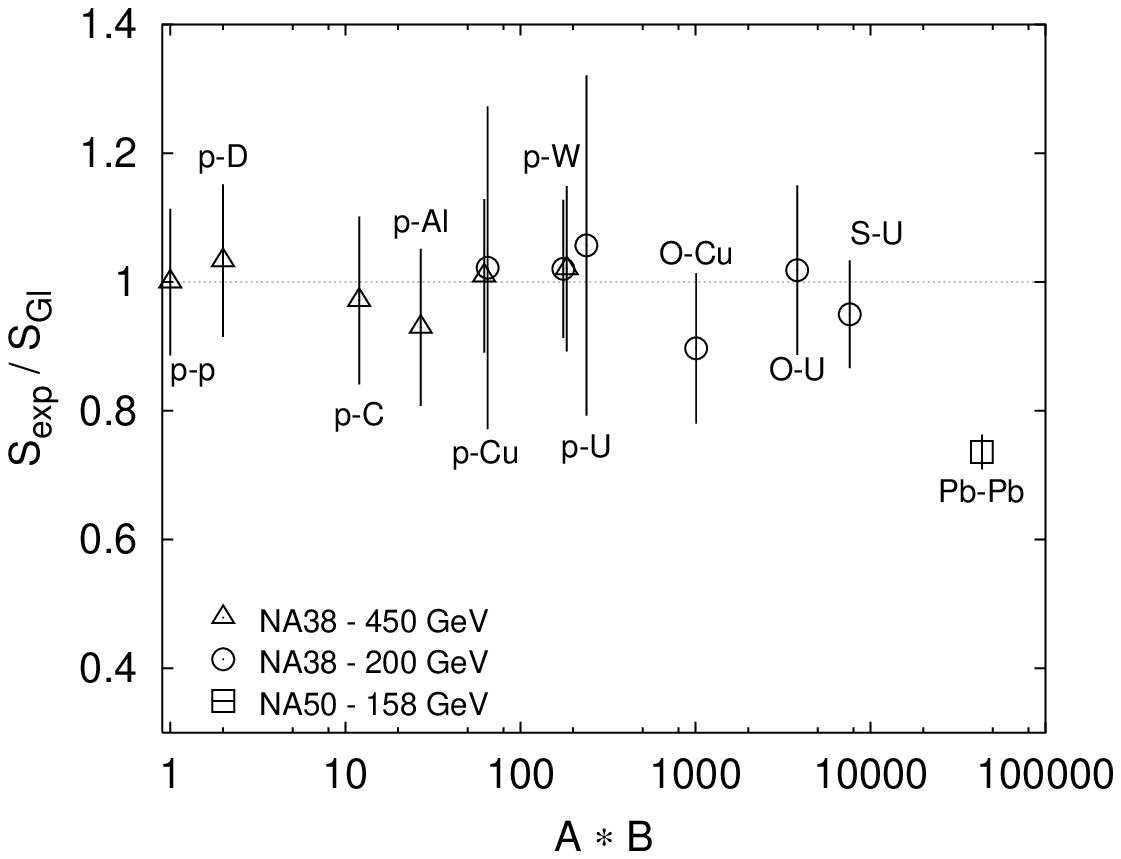,width=7.5cm,height=7cm}
\end{picture}\par
\caption{Centrality integrated rates for \J~production in $A-B$
collisions, normalized to pre-resonance absorption in nuclear matter
\cite{NA38-f,NA50}.}
\label{F5}
\end{minipage}
\end{figure}

\J~production in p-A collisions has been studied in fixed target
experiments with incident protons of momenta between 200 and 800 GeV/c
\cite{NA3}-\cite{NA38-f}. Data are taken mainly for $x_F > 0$, which
gives the nascent \J~momenta of 30 GeV/c or more in the target rest
frame. As a result, the transition $(\C\!-\!g) \to \j,~\x$ or \P~occurs
at the edge of the target nucleus or completely outside; the nuclear
medium of the target sees only the passage of pre-resonance states.
Since these have essentially the same size for all charmonia, the
observed attenuation of the production rates should be the same for
\J~as for \P. This is indeed found to be quite well satisfied, as shown
in Fig.\ \ref{F4} \cite{NA38-f}. Earlier attempts to explain
charmonium suppression in p-A interactions in terms of the
absorption of physical \J~states \cite{Gerschel} had encountered
difficulties precisely because of this feature. The equal attenuation
of \J~and \P~for $x_F>0$ is a natural consequence of pre-resonance
absorption; it can never be obtained for the physical \J~and \P~states
with their very different geometric sizes.

\par

The cross section for the absorption of pre-resonance charmonium in
nuclear matter can be determined through a Glauber analysis of p-A
data. Such an analysis of the most recent data for 200 and 450 GeV data
\cite{NA38-f} gives $\sigma_{\C g\!-\!N} = 6.9 \pm 1.1$ mb, in
accord with previous analyses \cite{KLNS} and theoretical estimates
\cite{KS6}.

\par

With this, the first two steps of the mentioned procedure for the
application of \J~suppression as deconfinement probe are completed. We
have a viable theoretical description of the elementary production
process, although some aspects of colour neutralization still remain
unclear \cite{polarize}. We know how this process is modified in
the confined medium provided by nuclear matter. The expected `normal'
behaviour is clear, and the probe can now be applied to nuclear
collisions in order to look for an onset of deconfinement. In Fig.\
\ref{F5}, the production rates measured in different p-A and
centrality-integrated A-B collisions with light ion beams
\cite{NA38-f} are seen to agree completely with `normal' pre-resonance
suppression; this agreement persists as well for the centrality
dependence of \J~production in S-U interactions, as determined by
measuring the associated transverse energy $E_T$  \cite{KLNS}. Hence up
to central S-U collisions, there is no unconventional sign.

Before turning to Pb-Pb collisions, we note a characteristic feature of
\J~suppression by deconfinement. The \J's measured in hadronic
collisions originate from three different sources: about 60\% are
directly produced \J's, while about 32\% arise from \X~decay and the
remaining 8\% from \P~decay. Since the decay widths are extremely small,
the decays occur far outside the interaction region; any produced
medium therefore effects (after the pre-resonance stage) the survival
of three different charmonium states. Dissociation calculations based
on a screened confining potential \cite{MTM} indicate that the \X~and
the \P~are dissociated essentially at the deconfinement point, while
the more tightly bound \J~state requires higher energy densities to
`melt'. This result is confirmed by collision calculations based on the
different binding energies of the three states \cite{KS3}; however,
here one finds that the \P, which is only 60 MeV below the open charm
threshold, can also be dissociated by any interaction in a confined
medium. These considerations predict a sequential suppression pattern
\cite{KS,GS}:\ first any medium, whether confined or deconfined, will
lead to \P~suppression. When the energy density is increased beyond the
deconfinement point $\e_c$, \X~suppression will set in, and beyond 1.5
- 2 $\e_c$, also the directly produced \J's will be suppressed.

\par

\J~suppression by colour deconfinement is thus predicted to have two
specific qualitative features: apart from the small contribution due to
\P~decay, it should set in quite suddenly, since we are dealing with
critical behaviour, and it should show a characteristic two-step
pattern, corresponding to sequential \X~and \J~suppression.

\par

\begin{figure}[tbp]
\setlength{\unitlength}{1cm}
\begin{minipage}[t]{7.0cm}
\begin{picture}(7.5,7.5)
\hspace*{-0.2cm}
\psfig{file=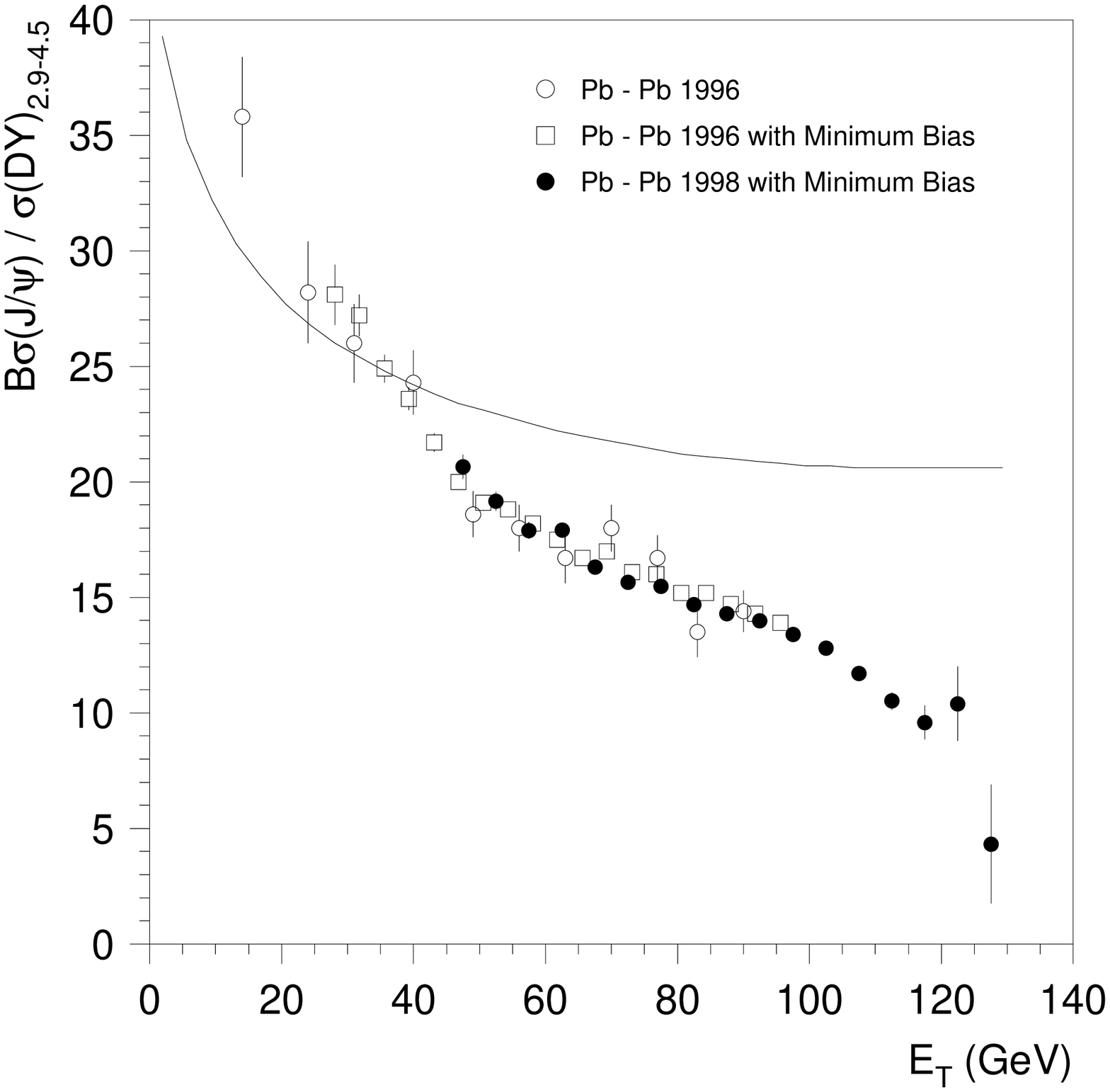,width=6.5cm,height=7cm}
\end{picture}\par
\caption{The $E_T$-dependence of the ratio \J/Drell-Yan production,
compared to the normal dependence due to pre-resonance suppression
\cite{NA38-f,NA50}.}
\label{F6}
\end{minipage}\hfill
\begin{minipage}[t]{7.0cm}
\begin{picture}(7.5,7.5)
\psfig{file=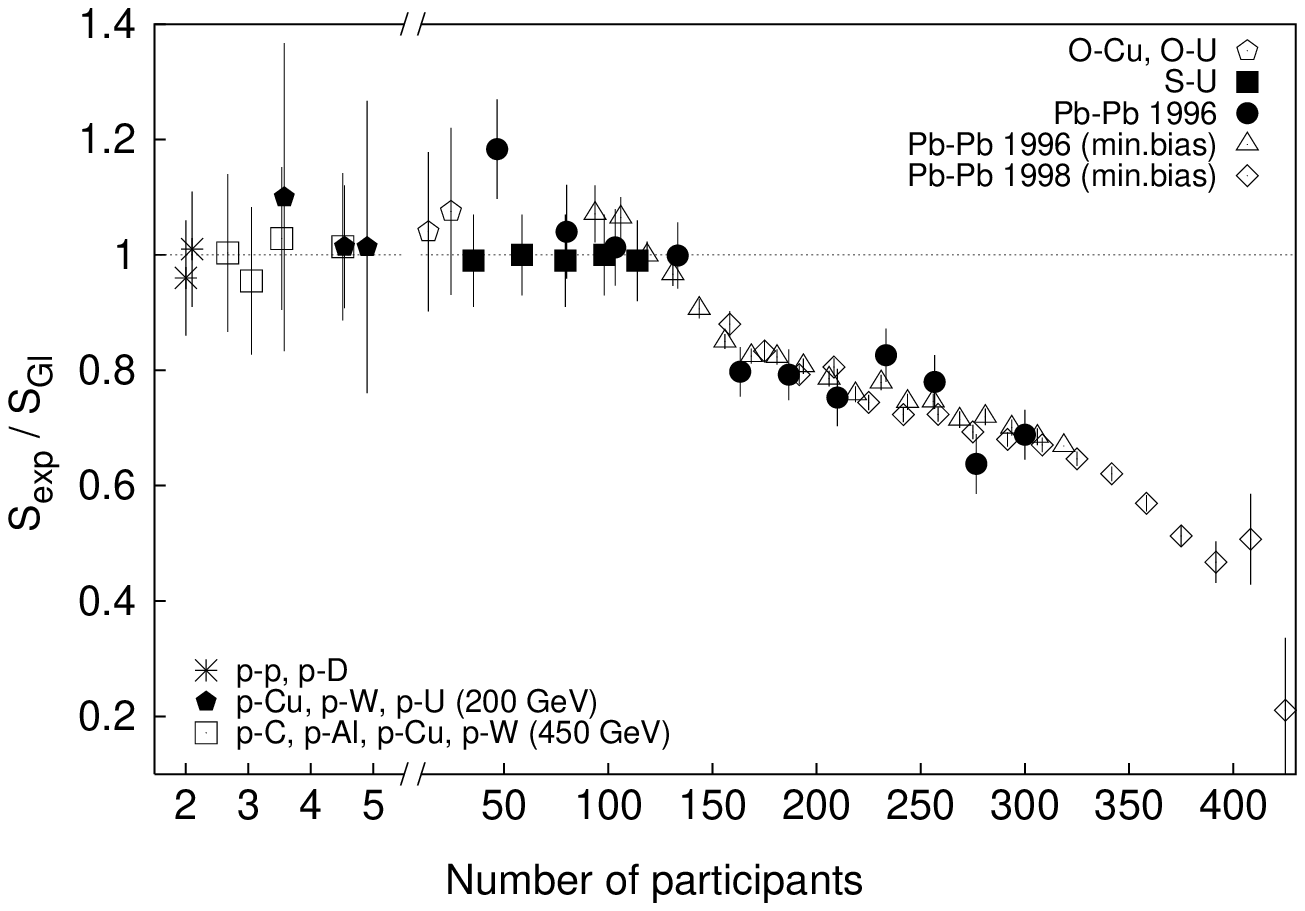,width=6.5cm,height=7cm}
\end{picture}\par
\caption{\J~production in A-B collisions, normalized to 
pre-resonance absorption in nuclear matter, as function
of the number of participating nucleons \cite{NA50,Nardi}.}
\label{F7}
\end{minipage}
\end{figure}

\begin{figure}[tbp]
\setlength{\unitlength}{1cm}
\begin{minipage}[t]{7.0cm}
\begin{picture}(7.5,7.5)
\psfig{file=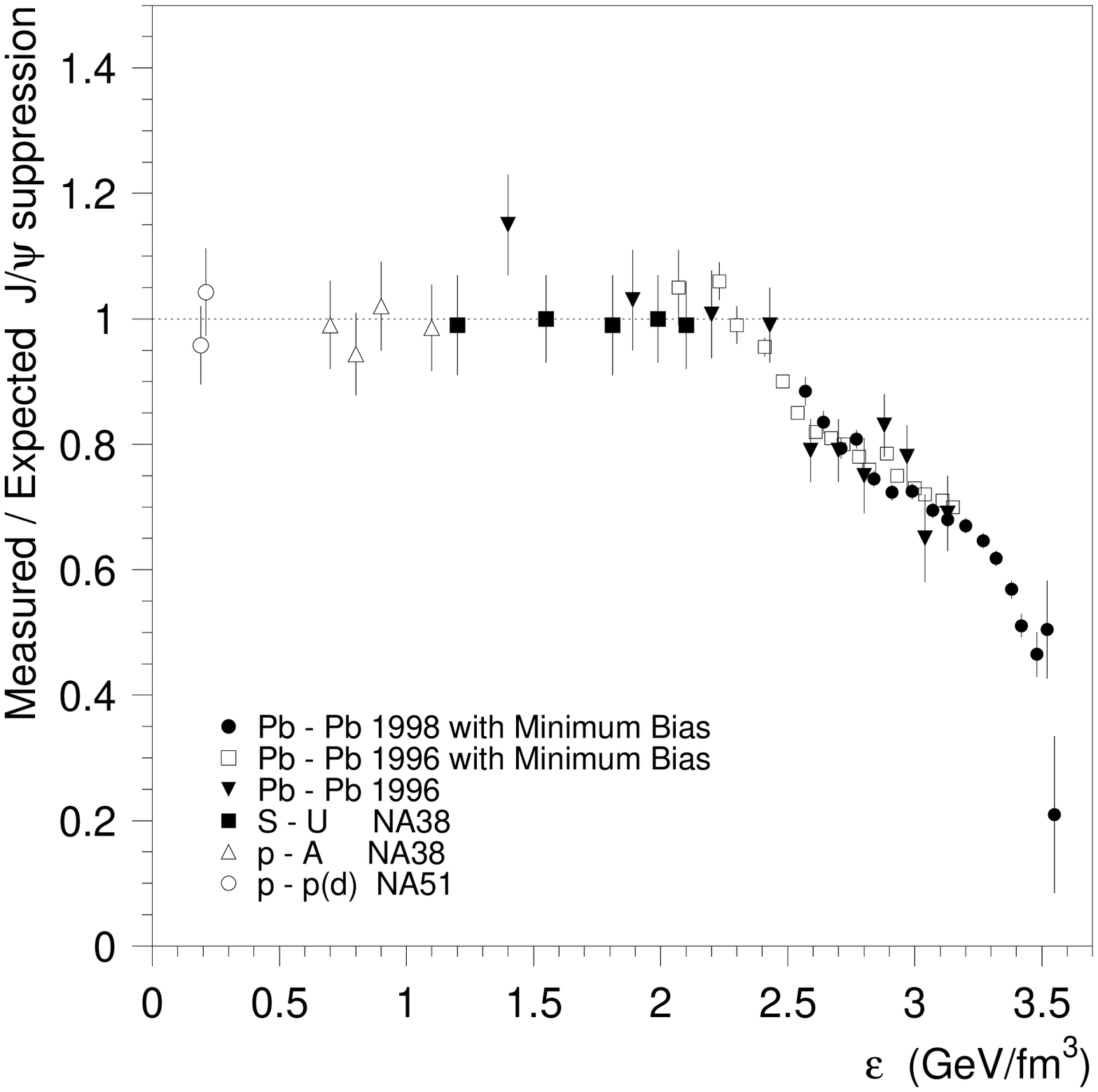,width=6.5cm,height=7cm}
\end{picture}\par
\caption{\J~production in $A-B$ collisions,
normalized to pre-resonance absorption in nuclear matter, as function
of the initial energy density \cite{NA50}.}
\label{F8}
\end{minipage}\hfill
\begin{minipage}[t]{7.0cm}
\begin{picture}(7.5,7.5)
\psfig{file=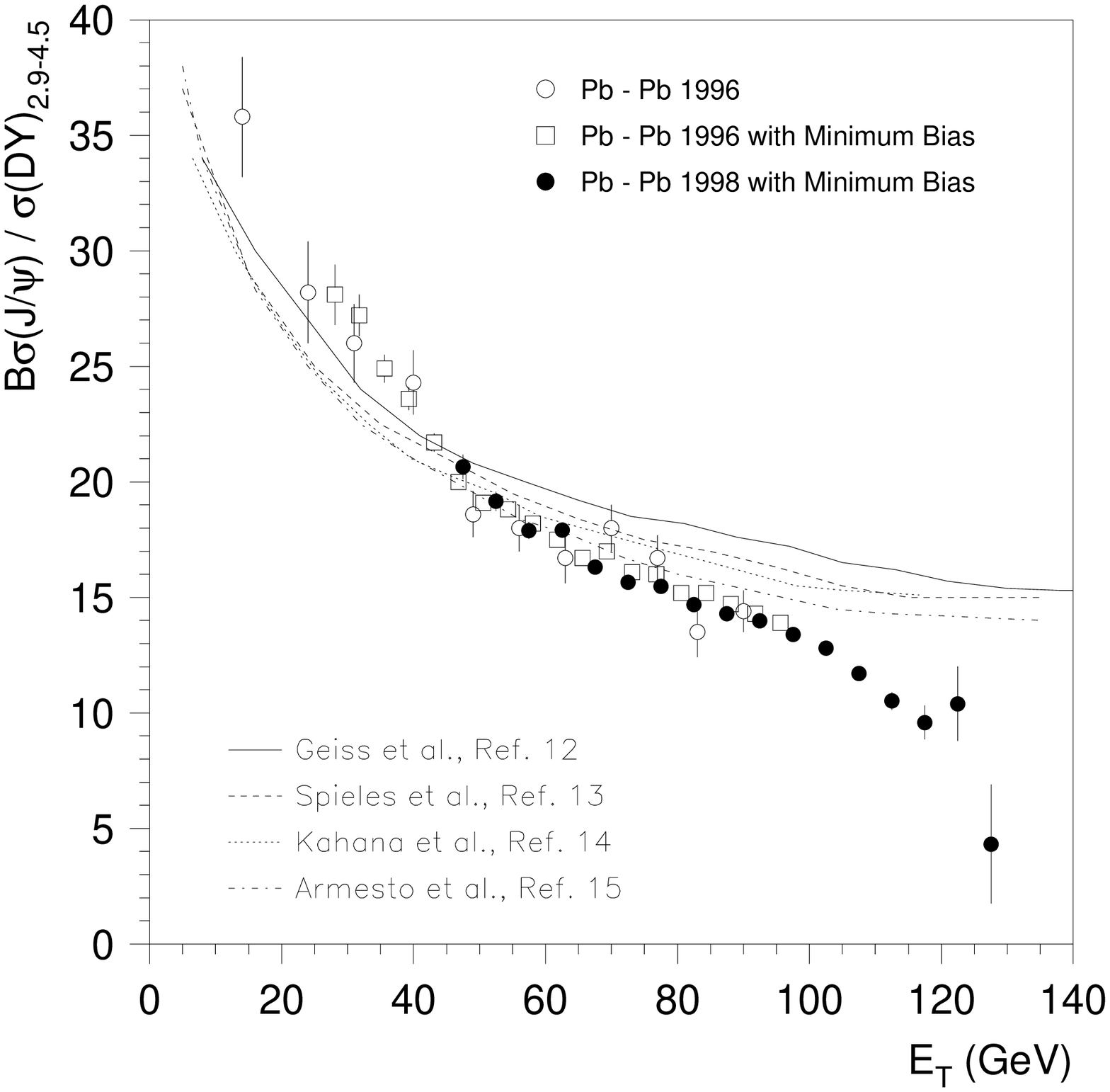,width=6.5cm,height=7cm}
\end{picture}\par
\caption{The $E_T$-dependence of the ratio \J/Drell-Yan production,
compared to different hadronic comover models, with references as given in
\cite{NA38-f}.}
\label{F9}
\end{minipage}
\end{figure}

The experimental studies over the past four years \cite{NA50}
have shown that in Pb-Pb interactions, \J~production is reduced beyond
the expected pre-resonance absorption (see Fig.\ \ref{F5}). In detail,
this anomalous suppression shows the following features:
\begin{itemize}
\vspace*{-0.2cm}
\item{In peripheral Pb-Pb collisions, \J~production follows the normal
pattern of pre-resonance suppression up to $E_T \simeq 40$ GeV.}
\vspace*{-0.2cm}
\item{With increasing centrality, the anomalous suppression sets in
quite suddenly at $E_T \simeq 40$ GeV, corresponding to $N_w \simeq
100$ participating nucleons.}
\vspace*{-0.2cm}
\item{At $E_T \simeq 100$ GeV, a second drop of \J~production occurs;
this corresponds to $N_w \simeq 300$ participating nucleons.}
\vspace*{-0.2cm}
\end{itemize}
\noindent
In Fig.\ \ref{F6}, the result is shown as function of the associated
transverse energy $E_T$, together with the expected form for
pre-resonance absorption. Using a zero-degree calorimeter, the
experiment also determines the number $N_w$ of participant nucleons for
each value of $E_T$, as noted above for the threshold points. In Fig.\
\ref{F7}, the \J~survival probability in the produced medium, i.e.,
the production rate divided by the pre-resonance absorption, is shown
as function of $N_w$, for all p-A and A-B data taken by the
collaboration. Using the free-streaming evolution model \cite{Bjorken},
one can convert $E_T$ into energy density for the different reactions.
This is shown in Fig.\ \ref{F8} \cite{NA50}, with the first
drop occurring just above 2 GeV/fm$^3$ and the second around 3
GeV/fm$^3$.

The observed two-step form of anomalous suppression is thus the one
predicted by deconfinement \cite{KS,GS}. In contrast, it is in
qualitative and quantitative disagreement wit all hadronic comover
models, as illustrated in Fig.\ \ref{F9}. We conclude that
the \J~suppression pattern observed by the NA50 collaboration indeed
constitutes an unconventional sign, which can be quite naturally
interpreted as the onset of deconfinement in nuclear collisions
\cite{NA50}.

\par

However, this conclusion is still accompanied by a number of open
questions which have to be answered in order to fully understand what
is happening.
\begin{itemize}
\vspace*{-0.2cm}
\item{What is the relevant variable which determines the onset of
deconfinement - is it indeed the energy density, as suggested in Fig.\
\ref{F8}? To answer this conclusively, experiments with different
nuclei are needed.}
\vspace*{-0.2cm}
\item{What is the origin of the rather steep final drop? The onset of
suppression for direct production does not reproduce it in the observed
strength \cite{HS-Torino}. Transverse energy fluctuations could
form another or additional mechanism \cite{Blaizot}.}
\vspace*{-0.2cm}
\item{How does the \P~fit into the picture? As expected, it shows the
same pre-resonance suppression as the \J. In all S-U collisions,
however, it is significantly suppressed, with a rather abrupt onset at
quite peripheral interactions. Because of the small binding energy, the
\P~could
 be dissociated by hadronic or partonic comovers.}
\vspace*{-0.2cm}
\item{How do the suppression patterns for the different charmonium
states vary with their transverse momentum? Here a normal and an
anomalous behaviour can also be defined \cite{KNS}; present data
are still inconclusive.}
\vspace*{-0.2cm}
\item{So far, charmonium production was studied experimentally with
Drell-Yan production as reference. In view of the observed suppression
of bound $\C$ states in nuclear collisions, the unsuppressed behaviour
($\sim A^2$) for unbound (open charm) $\C$ production should be
verified. In any case, open charm production provides a better
reference, since it involves the same gluon distribution functions as
charmonium production.}
\vspace*{-0.2cm}
\end{itemize}
\noindent
Some of these questions could well still be addressed by the SPS. The
scaling variable, as noted, can be clarified by A-A experiments at lower
A than Pb. To identify the origin of the second drop requires either
experiments at lower A (SPS) or at higher incident energy (RHIC): does
the drop shift to lower $N_w$ or higher $E_T$, respectively, as
predicted if transverse energy fluctuations are responsible? In any
case, a confirmation of the observed \J~suppression pattern by RHIC,
perhaps with even more pronounced thresholds, seems of basic importance
for the entire QGP study. Finally we note that a recent study \cite{Thews}
invokes final state reinteractions to predict for A-A collisions
(normalized to A) at very high energies a \J~enhancement rather
than suppression relative to p-p production rates. Similar conclusions
are obtained through thermal colour neutralisation \cite{PBM}, although
this model is not compatible with preliminary data on \P/\J~production
in Pb-Pb, compared to S-U \cite{psi',tsukuba}. The first \J~data from
RHIC, even if of low statistics, will thus already provide some
interesting conclusions.

\bigskip

\noindent
{\bf 3.\ In-Medium Hadron Modifications}

\bigskip

High energy nuclear collisions eventually produce a hadronic state, and
it is clearly of interest to check if these hadrons form at some stage
a dense, interacting medium. The \P~suppression observed in S-U
collisions could well be an indication for this, since the produced
medium here does not effect \J~production. A direct probe to study this
question is provided by dilepton spectra in the hadronic resonance mass
range. The decay width of the $\rho$ into a lepton pair gives this
resonance a life-time of about 1 fm, so that any in-medium modifications
should show up in the measurable dilepton spectrum \cite{Shuryak,K-M}.
Most striking would be mass changes in the vicinity of the chiral
symmmetry restoration point \cite{Brown}; it is clear, however, that
an interacting medium as such can quite generally lead to modified
resonance properties \cite{Wambach}.

\par

\begin{figure}[htb]
\centerline{\epsfig{file=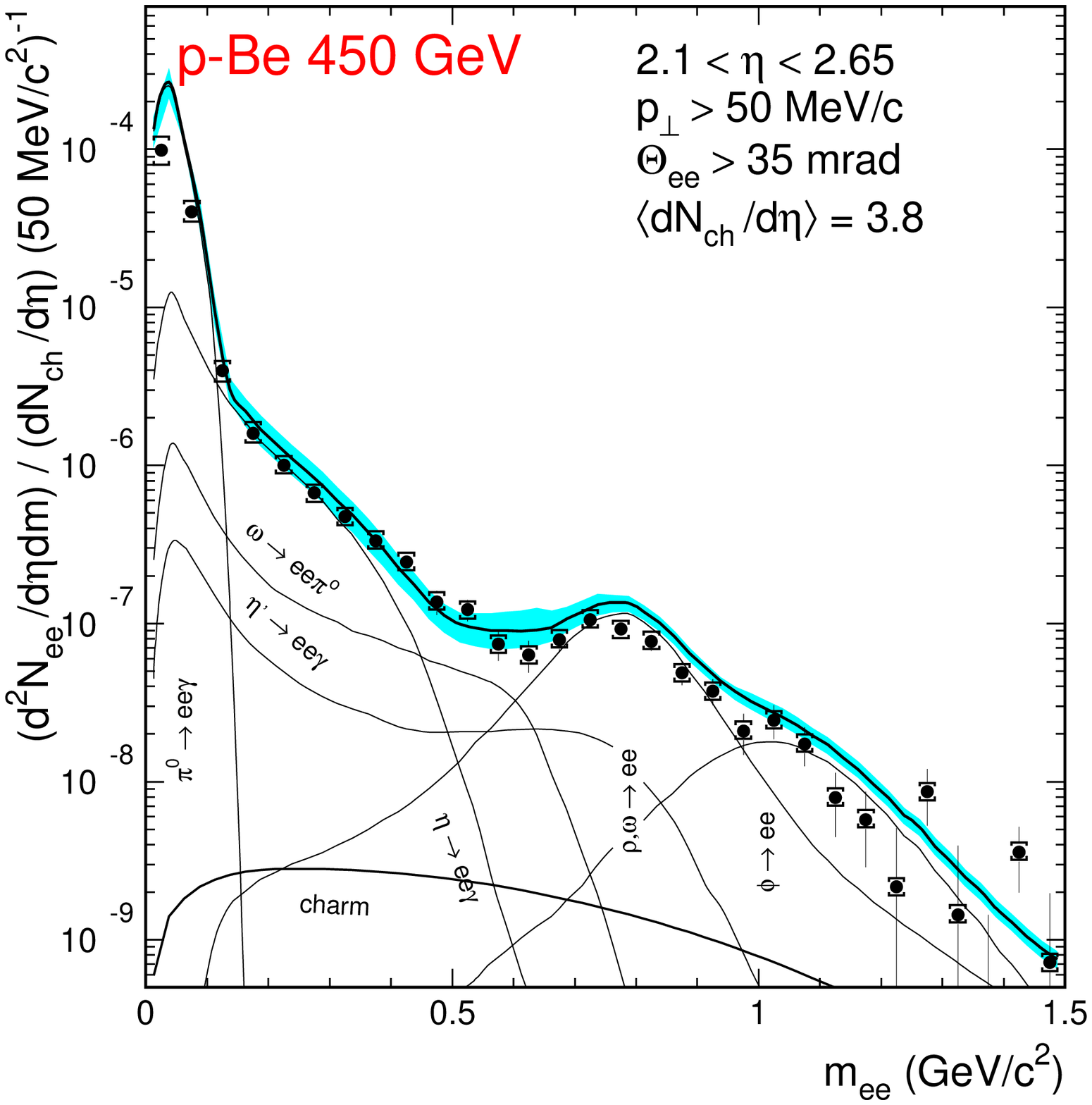,height=70mm,width=60mm}
\hspace*{2cm}
\epsfig{file=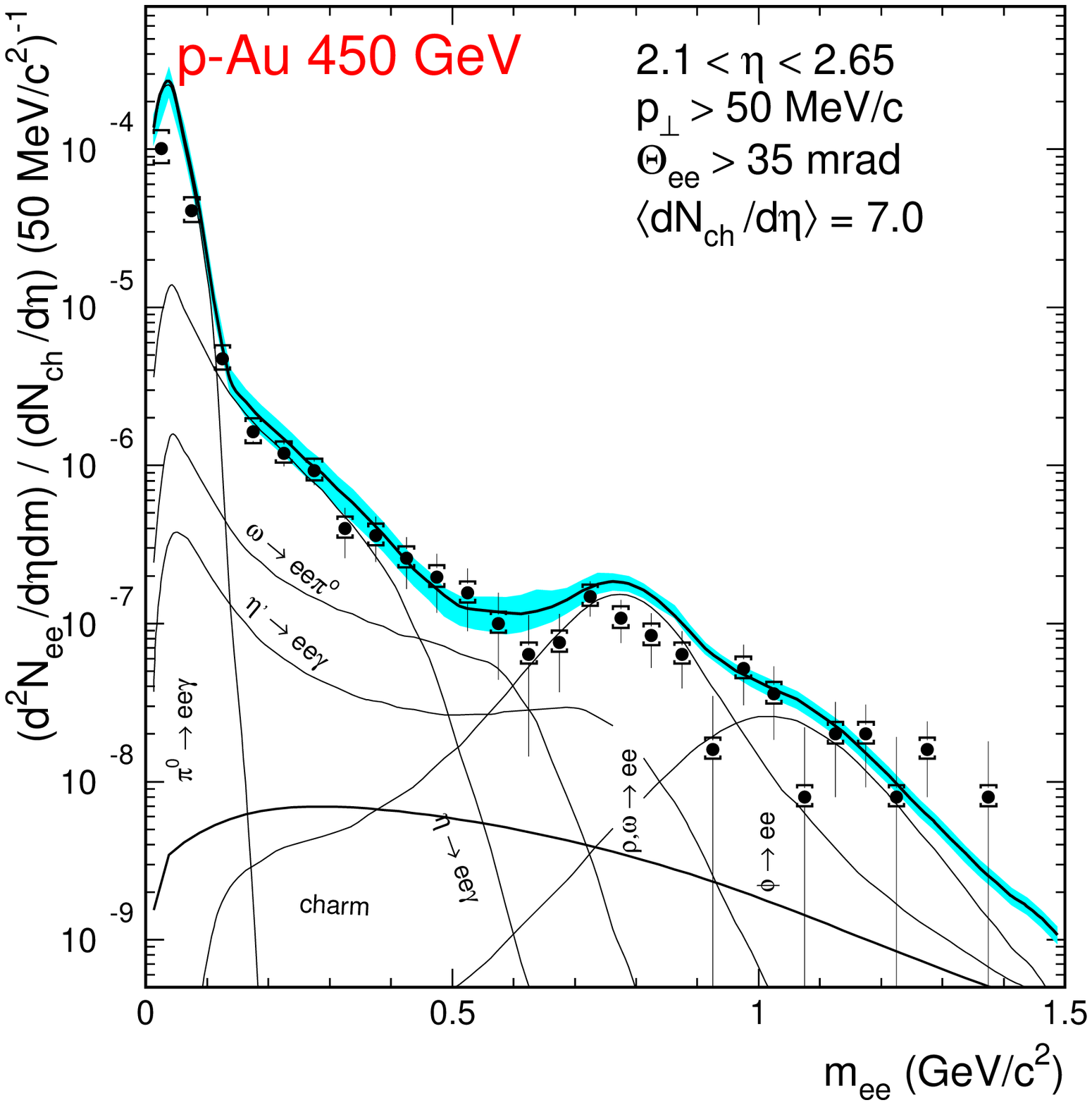,height=70mm,width=60mm}}
\hspace*{-1cm}
\vspace*{-1cm}
\caption{The dilepton spectrum in p-Be (left) and p-Au (right) collisions,
compared to the combined results (band) of hadronic decay contributions 
\cite{Ceres}.}
\label{F10}
\end{figure}

\begin{figure}[htb]
\centerline{\psfig{file=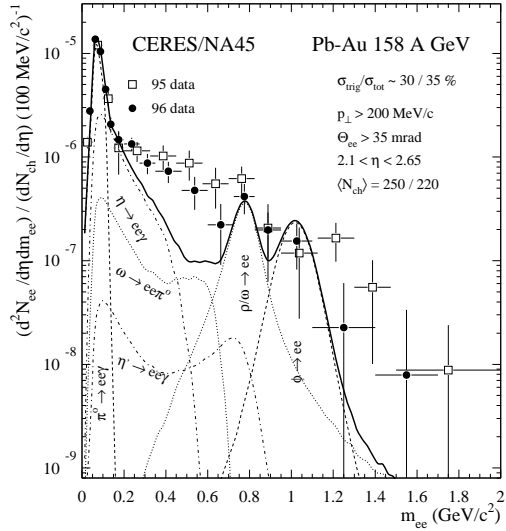,width=7cm}}
\caption{The dilepton spectrum in Pb-Pb collisions,
compared to the combined results (band) of hadronic decay contributions 
\cite{Ceres}.}
\label{F11}
\end{figure}

The dilepton spectrum from very low masses up to the \J~was measured by
several SPS experiments \cite{Kluberg,Ceres,NA38/50}.
In Fig.\ \ref{F10} it is seen that the spectrum measured for proton
beams incident on both light and heavy targets is accounted for quite
well by the decay of the known and observed hadronic states decaying
into lepton pairs. For S-U and Pb-Pb collisions this is no longer the
case; an excess is observed both for low mass pairs, below the
$\rho-\omega$ peak, and for intermediate mass pairs, between $\phi$ and
\J~(see Fig.\ \ref{F11}).

The low mass enhancement has been the subject of much theoretical
attention. A decrease of the $\rho$ mass in a hot medium near chiral
symmetry restoration could indeed lead to such an effect \cite{Brown};
however, lattice studies do not show any evidence of a changed $rho$
mass over a significant temperature range. On the other hand, in-medium
broadening of the resonance due to hadronic interactions can as well
reproduce the observed enhancement \cite{Wambach}. Hence it is at
present not clear if the observed modifications of the dilepton spectrum
indeed constitute an unconventional sign. The rather low
signal-to-background ratio of present data moreover does not allow a
very quantitative study of the effect. It is thus not clear yet if there
is really a low mass enhancement and a distinct separate enhancement at
higher masses, or if the two observations are one effect. So far, there
is also no evidence of any onset or threshold. Some interesting recent
work \cite{Rapp} has tried to interpret in particular the higher mass
enhancement in terms of thermal dileptons. In summary, here more
precise experimental data seem very necessary before any conclusion can
be reached.

\bigskip

\noindent
{\bf 4.\ Strangeness and Thermalization}

\bigskip

\begin{figure}[htb]
\centerline{\psfig{file=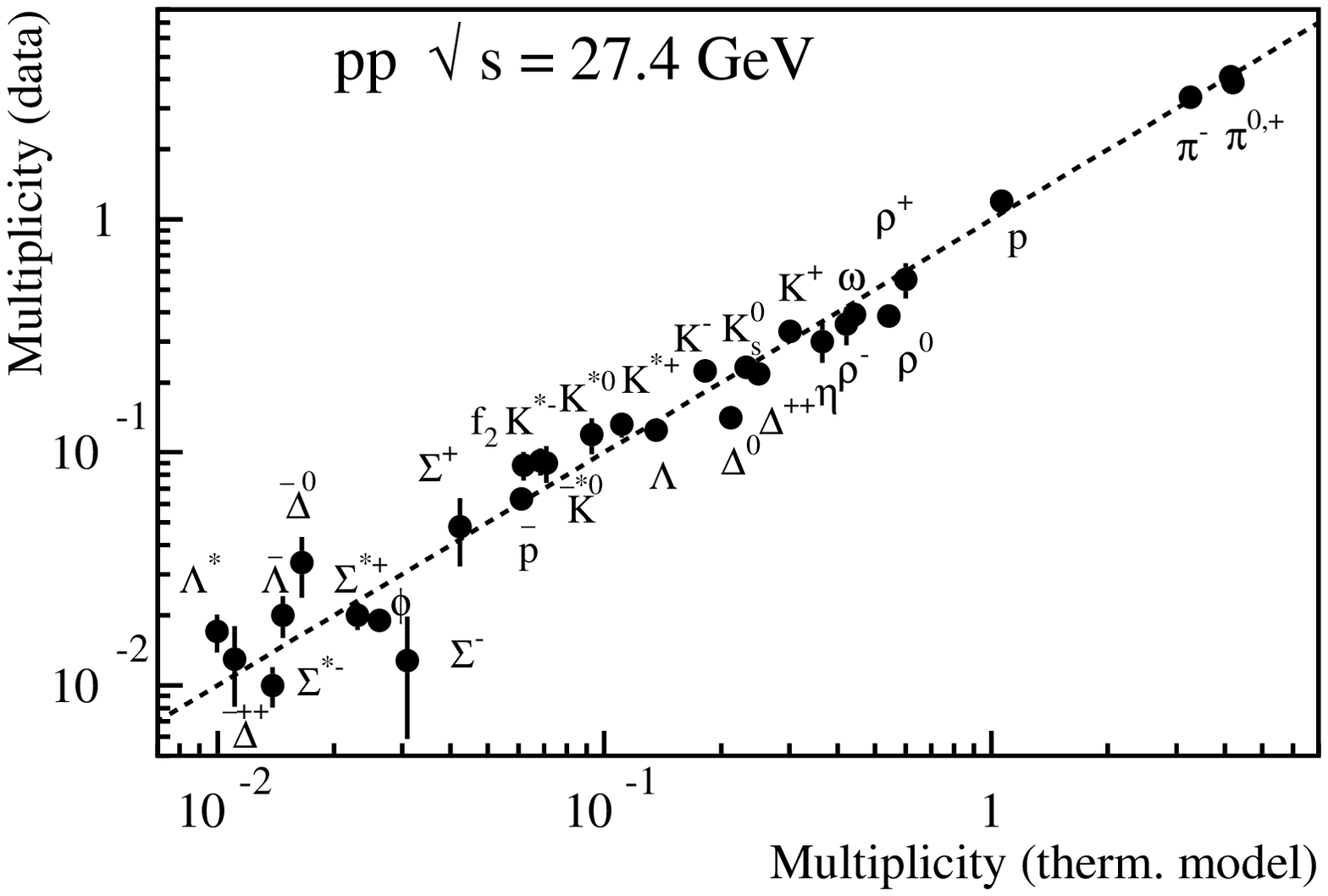,width=8.5cm,height=7cm}}
\caption{Hadron abundances for p-p collisions, compared to the prediction
of an ideal resonance gas at $T_h=170$ MeV and $\gamma_s=0.5$
\cite{Becattini}.}
\label{F12}
\end{figure}

One of the most remarkable observations in high energy multihadron
production by whatever means is that the relative abundances of the
secondary hadrons seem to be quite well described by an ideal resonance
gas of a temperature $T_h \simeq$ 170 MeV \cite{Hagedorn,Becattini}.
The only and systematic exception is due to strange particles: in p-p,
p-$\bar{\rm p}$ and $e^+e^-$ collisions, these are produced less
abundantly than expected in a resonance gas scheme. This reduction
can be accounted for by a general strangeness suppression factor
$\gamma_s \simeq 1/2$: for each strange or antistrange quark in the
hadron in question, one power of $\gamma_s$ reduces the production
abundance. Thus the entire relative production spectrum of sometimes up
to 30 different resonances can be described in terms of two parameters,
$T_h \simeq 170$ MeV and $\gamma_s \simeq 1/2$ \cite{Becattini}, as
illustrated in Fig.\ \ref{F12}.

\begin{figure}[htp]
\vspace*{-1cm}
\centerline{\epsfig{file=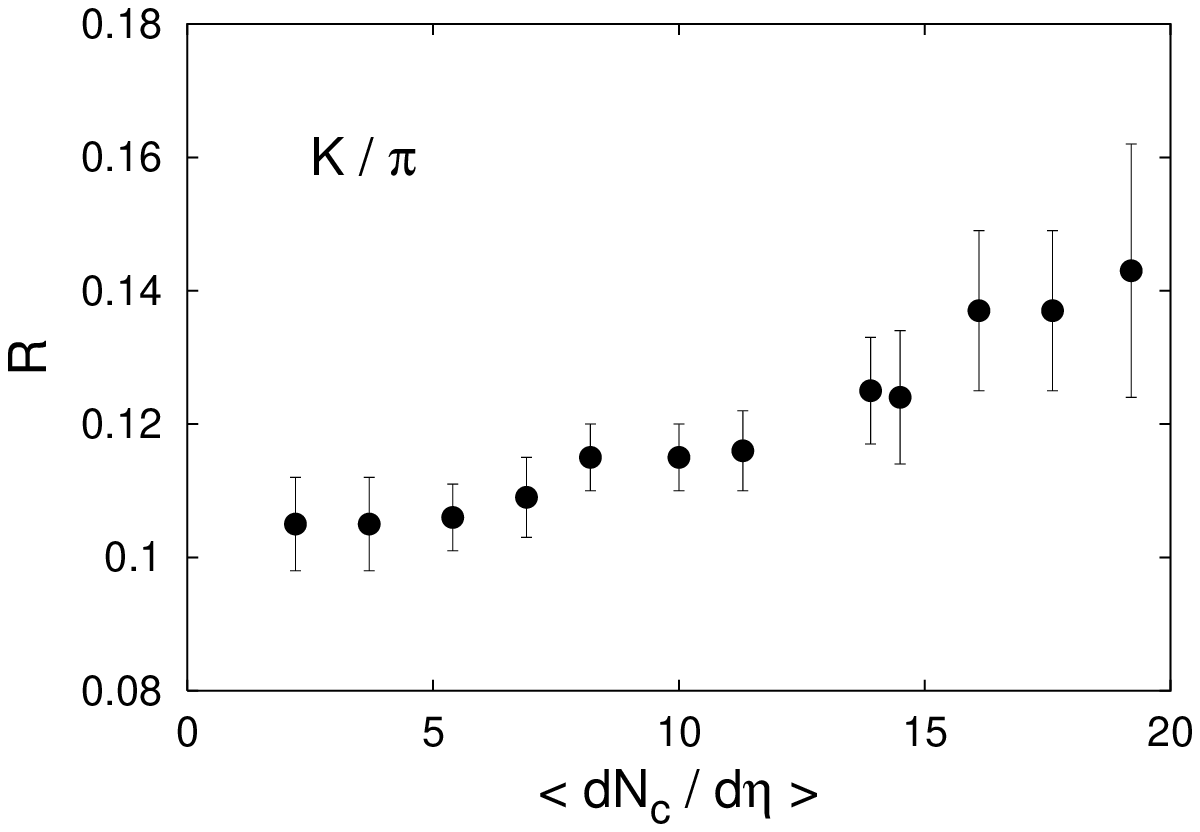,height=70mm,width=70mm}
\hspace*{2.5cm}
\epsfig{file=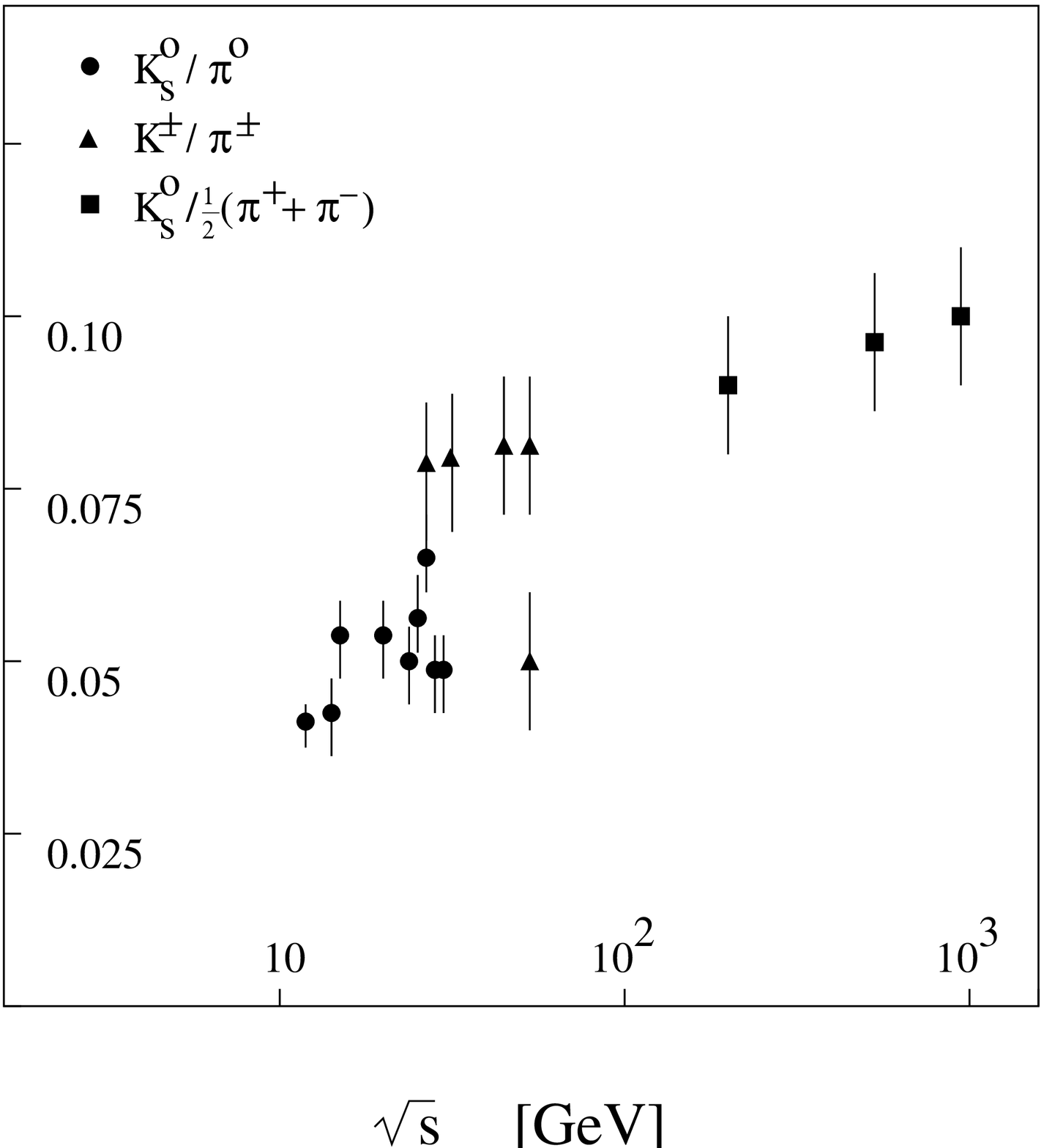,height=60mm,width=60mm}}
\hspace*{-0.5cm}
\caption{The ratio of kaon to pion production in p-${\bar {\rm p}}$
collisions (left)
as function of the associated multiplicity \cite{Alexo} and (right)
as function of the incident energy \cite{Ansorge}.}
\label{F13}
\end{figure}

The origin of such a ``thermal" production scheme and of the observed
strangeness suppression is not at all clear. Since the description is
even valid in $e^+e^-$ interactions with very few hadrons per unit
rapidity, it can definitely not be caused by final state hadron
interactions. Conceptually the most convincing picture is that any high
energy collision implies a passage of colour charges through the
vacuum, which is excited by this passage and subsequently neutralizes
the excitation by a statistically governed formation of
quark-antiquark pairs and gluons. As the bubble
of excited vacuum expands, the quarks and gluons have to combine to
produce the observed secondary hadrons, which reflect their
``statistical" origin and the universal hadronization temperature $T_h$
\cite{Dokshitzer}. While the hadronization temperature is universal, the
energy density of the initial excited vacuum bubble will increase with
increasing collision energy, and it will presumably also be higher in
A-A than in p-p collisions at the same incident beam energy. A possible
cause of the observed strangeness suppression is thus the higher
strange quark mass, which makes it more difficult to form $s{\bar s}$
pairs than pairs of light $u$ or $d$ quarks. For an ideal quark gas, one
would expect
\be
\gamma_s = 1 - {m_s^2 \over 2 T_0^2},
\label{4.1}
\ee
where $T_0 \geq T_h$ is the initial temperature of the excited vacuum.
An immediate prediction of this scheme is a diminishing of
strangeness suppression in p-p (or p-$ {\bar {\rm p}}$) collisions,
either with increasing incident energy or at fixed energy with the
number of produced secondaries per unit rapidity (i.e., with the
energy deposit due to the collision). In Fig.\ \ref{F13},
this is indeed observed, since in both cases the ratio of kaons to
pions increases considerably, although the hadronization temperature
$T_h$ determined through multiplicity integrated abundances remains
the same.

\par

\begin{figure}[htp]
\vspace*{-1.5cm}
\setlength{\unitlength}{1cm}
\begin{minipage}[t]{7cm}
\begin{picture}(7.5,7.5)
\hspace*{-0.2cm}
\vspace*{-1cm}
\psfig{file=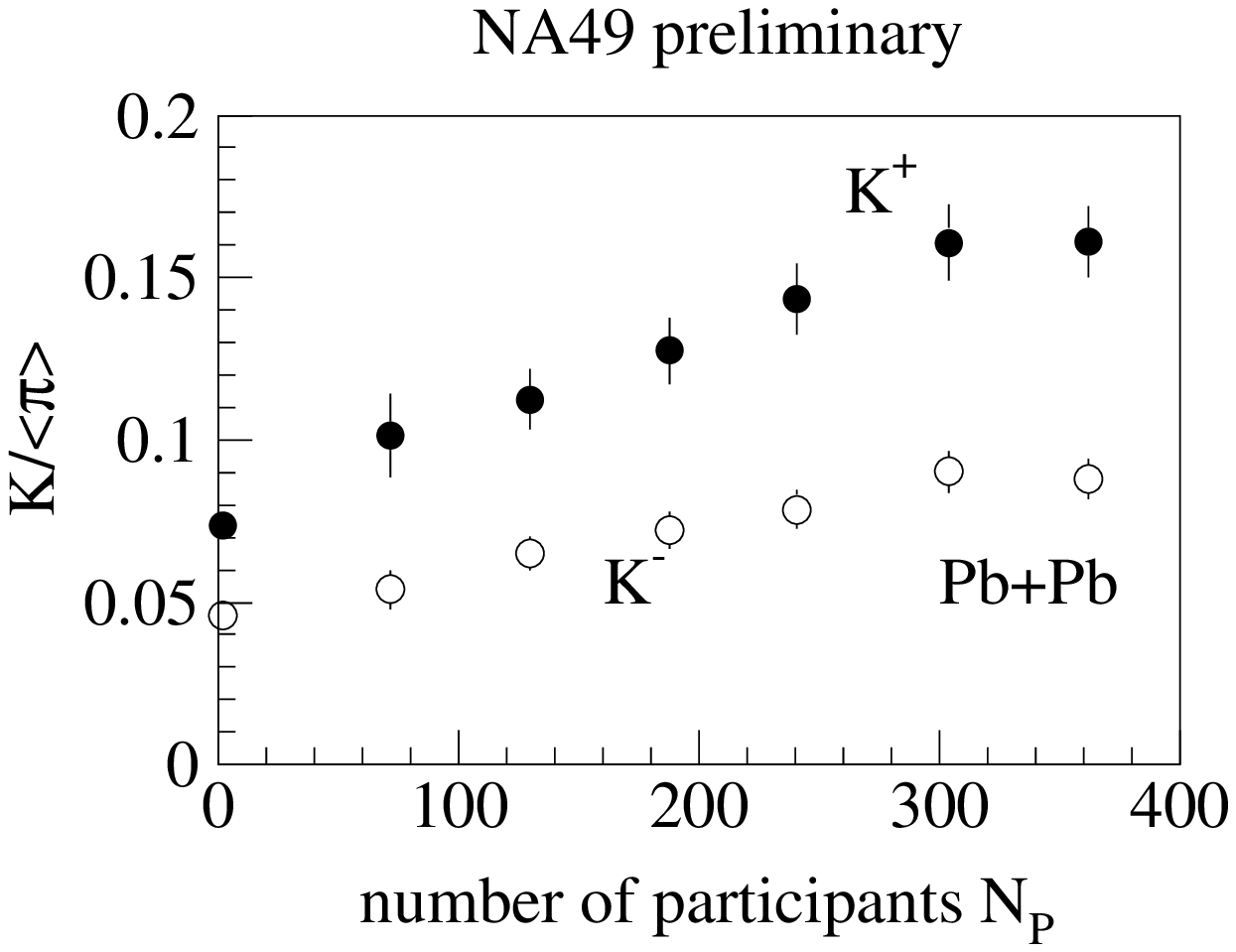,width=7cm,height=6cm}
\end{picture}\par
\caption{The ratio of kaon to pion production in Pb-Pb collisions 
\cite{Hoehne}.}
\label{F14a}
\end{minipage}\hfill
\begin{minipage}[t]{7cm}
\begin{picture}(7.5,7.5)
\hspace*{-0.4cm}
\psfig{file=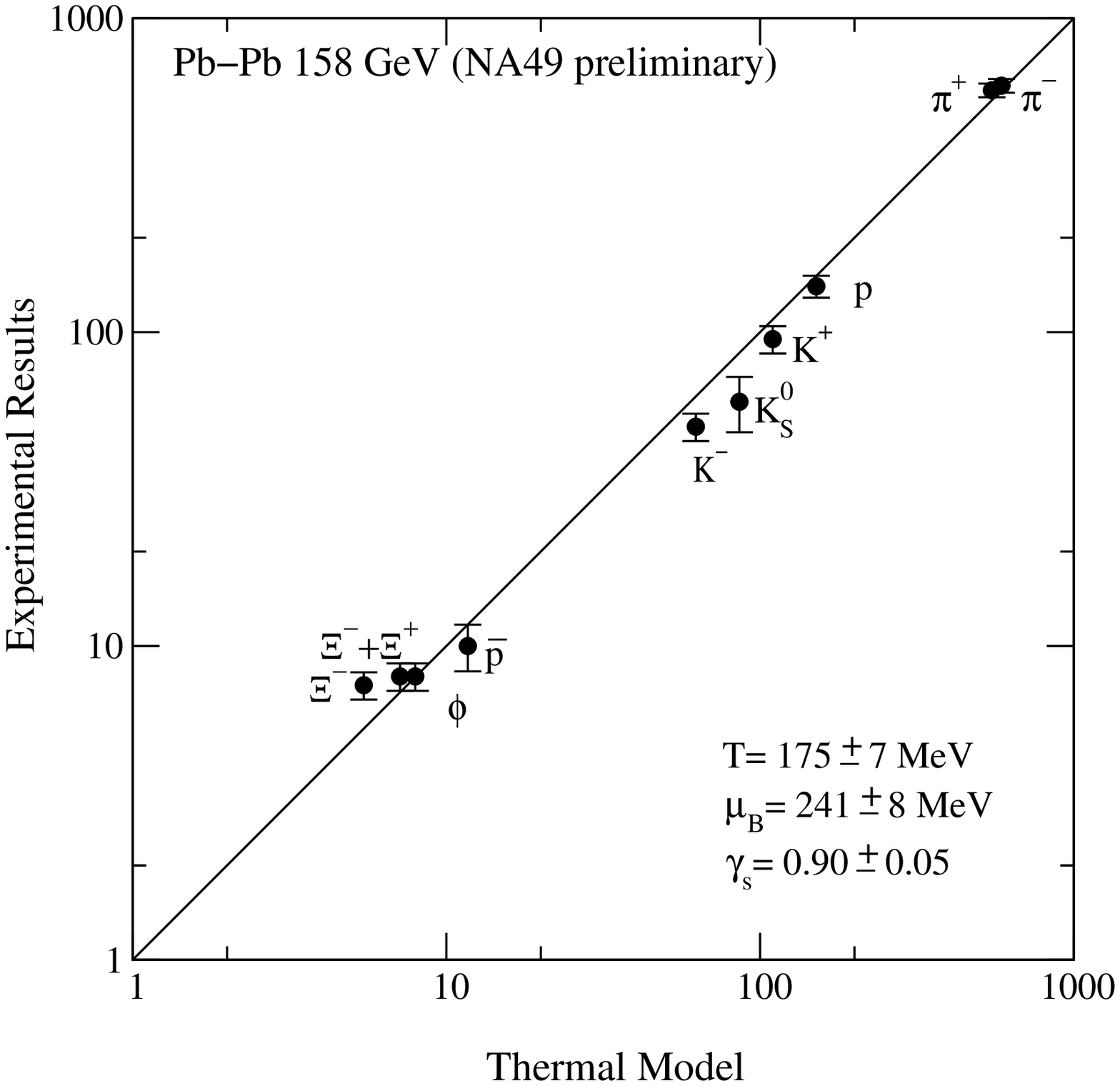,width=8.2cm,height=8cm}
\end{picture}\par
\caption{Hadron abundances for Pb-Pb collisions, compared to the prediction
of an ideal resonance gas \cite{Cleymans}.}
\label{F14}
\end{minipage}
\end{figure}

For nucleus-nucleus collisions, a strangeness enhancement from
the hadronic $\gamma_s \simeq 1/2$ towards $\gamma_s =1$ was predicted
as consequence of colour deconfinement \cite{RM-82,R-82}. In a
sufficiently hot QGP, the higher strange quark mass would become less
and less important, so that $u,d$ and $s$ quarks should eventually
appear in equal abundance. If the cooling of the QGP prodeeded rapidly
enough to preserve this relative abundance, the final hadronic state
obtained from a hadronized QGP in nucleus-nucleus collisions should
contain more strange hadrons than that formed in p-p interactions, i.e,
without an initial QGP. In Fig.\ \ref{F14a} we see that Pb-Pb collisions
at the SPS indeed lead to enhanced strangeness production, as again
reflected in an increasing kaon to pion ratio. In Fig.\ \ref{F14},
it is seen that the resonance gas description remains valid for
centrality-integrated abundances, with the same $T_h$ as above, but
with $\gamma_s$ close to unity.

Another possible origin for the strangeness suppression observed in
$e^+e^-$, p-p and p-${\bar {\rm p}}$ interactions could be the
relatively small number of strange particles produced. The ideal
resonance gas in its usual grand canonical formulation includes
strangeness conservation on the average. If only one kaon and its
antiparticle are produced in the event, strangeness conservation must
be imposed exactly, and this leads to a strong appearent suppression
compared to grand canonical rates \cite{Hagedorn-GC,Kred}. To
illustrate: in the grand canonical formulation, a kaon acquires a
Boltzmann factor $\exp \{-m_K/T_h\}$, while exact strangeness
conservation reduces this to $\exp \{-2m_K/T_h\}$, since the production
of one kaon requires that of one antikaon (we neglect associated
production for the sake of argument). These considerations lead to a
strangeness suppression which is strong at low relative multiplicities
and vanishes for large multiplicities when the grand canonical limit is
reached \cite{Kred}. The resulting suppression form is quite similiar
to the pattern observed in the production of strange baryons and
antibaryons, ranging from ``suppression" in p-p and p-A to an
``enhancement" in nuclear collisions \cite{Kred}.

\par

In summary: since strangeness enhancement is observed already in
hadron-hadron collisions for increasing energy or associated
multiplicity, it is {\sl a priori} not related to any macroscopic or
bulk effects, much less to colour deconfinement. Vacuum excitation as
well as exact strangeness conservation provide conventional monotonic
descriptions of the enhancement observed in high energy nucleus-nucleus
compared to p-p collisions. Before applying strangeness as a probe for
deconfinement, it is therefore necessary to define a normal strangeness
pattern, as seen in hadronic collisions and hence not related to QGP
formation. This given, one could then check if sufficiently central
nuclear collisions lead to any additional, non-montonic and hence
unconventional effects.

\par

\begin{figure}[htb]
\hspace*{-0.5cm}
\centerline{\epsfig{file=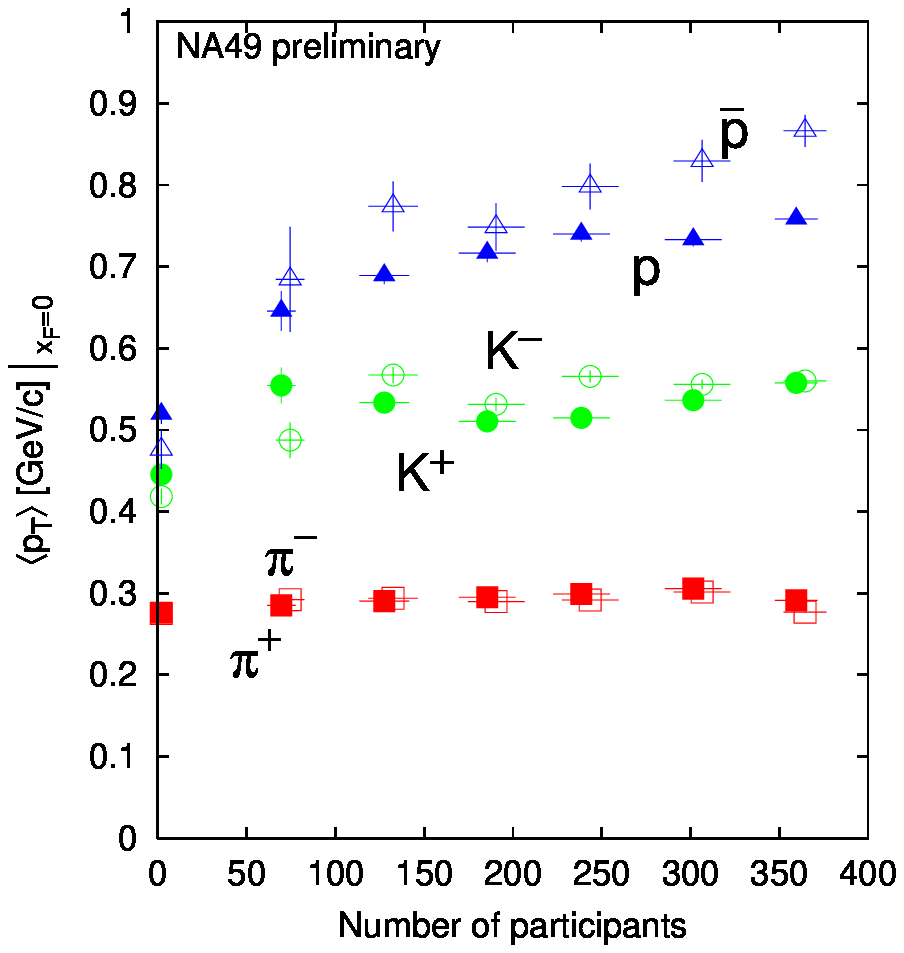,height=66mm,width=55mm}
\hspace*{2cm}
\epsfig{file=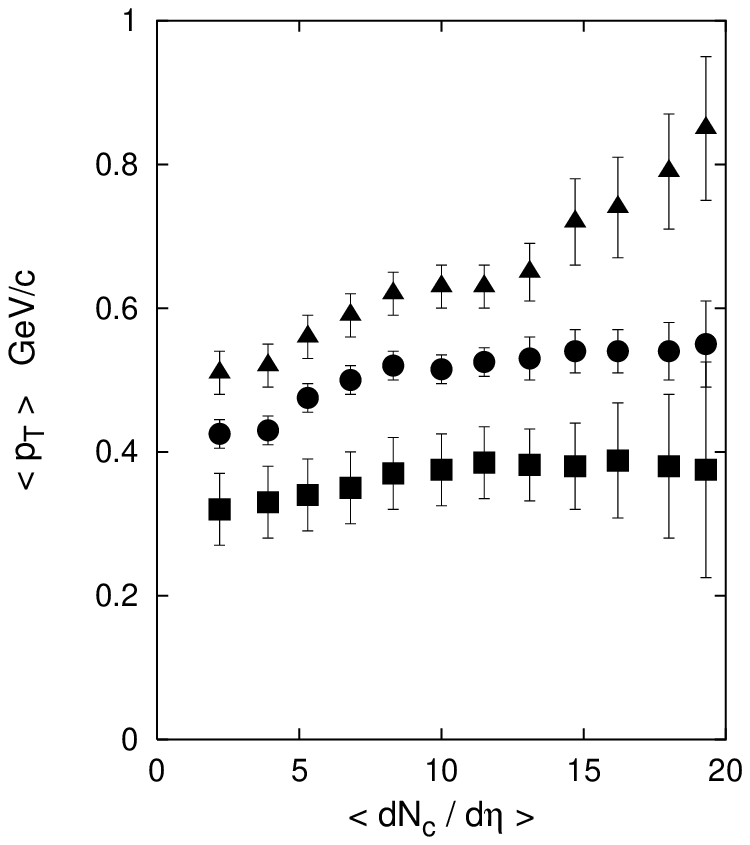,height=65mm,width=55mm}}
\hspace*{-1.5cm}
\caption{The average transverse momentum of different secondaries (left) in
Pb-Pb \cite{Stock} and (right) p-${\bar {\rm p}}$ collisions, 
same symbols \cite{Alexo}.}
\label{F15}
\end{figure}

The situation concerning transverse momentum broadening is quite
similar. In nuclear collisions, expansion and flow of the produced
medium are expected to lead to a broadening of the transverse momentum
spectrum of secondary hadrons, and it has been suggested that the
form of the broadening might reflect the transition to a QGP
\cite{VH}, flow effects \cite{flow} or the initial state scattering
in the medium \cite{initial}. Indeed a
mass-dependent broadening of $p_T$-spectra is observed e.g.\ in Pb-Pb
collisions; it increases with centrality (see Fig.\ \ref{F15}). However,
a very similar pattern, also increasing with particle mass, is observed
in p-${\bar{\rm p}}$ collisions \cite{Alexo}; it increases with the
associated multiplicity, as also shown in Fig.\ \ref{F15}.

In addition, there are other broadening effects known from p-p and p-A
collisions. In hadron-hadron interactions, there is the so-called
``seagull" effect, due to emission from a scattered primary nucleon; it
leads to a increase of $p_T$ with $x_F$, until the kinematic limit
requires a decrease. In p-A interactions,
there is the Cronin effect \cite{Cronin}, due to initial state
scattering of the nucleon on its passage through the nucleus.
Thus here as well the challenge is to define normal broadening, not
related to deconfinement, as observed in p-p and p-A collisions, and
then check if nuclear collisions show more than this and if there is any
kind of onset of anomalous behaviour. So far, this does not seem to be
the case.

\bigskip

\noindent{\bf 5.\ Catching the Elusive}

\bigskip

Looking back to the beginning of the search, we note a definite
progress in method: it is now clear how a systematic investigation
must be carried out. Whatever probe is to be used, it must first be
gauged in the absence of bulk matter, i.e., in p-p and/or p-${\bar{\rm
p}}$ collisions. Following this, its behaviour must be gauged in normal
nuclear matter as cold, confined medium. If possible, it should then be
tested in a hot confined medium, e.g., in the collision of light
nuclei. When all these tests clearly define the conventional behaviour
of the probe, we can start to look for an onset of anomalous behaviour.

\par

Where are we in this program with the different probes enumerated at the
beginning of this appraisal?
\begin{itemize}
\vspace*{-0.2cm}
\item{For {\bf charmonium suppression}, normal behaviour is well-defined
through p-A collisions and understood in terms of pre-resonance
absorption in nuclear matter. This also accounts for all collisions up
to central S-U interactions. In Pb-Pb collisions, there is a clear
onset of anomalous behaviour once the collisions reach a certain
centrality. The anomalour suppression pattern is in accord with the
deconfinement predictions of stepwise $\x_c$ and direct \J~suppression,
while it is in complete contradiction to all conventional
hadronic scenarios.}
\vspace*{-0.2cm}
\end{itemize}
\noindent
 \J~suppression therefore constitutes the first and so far only observed
unconventional sign. For its full understanding, however, a number
of open questions still have to be answered.
\begin{itemize}
\vspace*{-0.2cm}
\item{The {\bf dilepton enhancement} in nucleus-nucleus collisions
is specified in terms of a normal
behaviour defined through hadronic decays in p-p and p-A collisions.
In all nucleus-nucleus collisions studied, there are deviations; so far,
no clear onset for these is observed. A conventional theoretical
interpretation, based on resonance broadening in an interacting hadronic
medium, appears to reproduce the effect. Interesting alternative
explanations invoke changing hadron masses or thermal dileptons.}
\vspace*{-0.2cm}
\end{itemize}
\noindent
For more definite conclusions, also concerning the relation of low
and intermediate mass dilepton enhancement, more precise data are
necessary.
\begin{itemize}
\vspace*{-0.2cm}
\item{A {\bf strangeness enhancement} for increasing multiplicity or
centrality is observed in all hadronic collisions, from p-p to A-A.
There is so far no normal form and no observed onset. The effect may
well provide for all cases a thermometer for the initial partonic
environment; an increase of relative strangeness production between p-p
and A-A is also predicted if strangeness conservation is taken into
account exactly.}
\vspace*{-0.2cm}
\item{A {\bf transverse momentum broadening} is also observed quite
generally in hadronic interactions, increasing with multiplicity in p-p,
with A in p-A, and with centrality for A-A collisions. There are various
normal mechanisms, such as initial and final state scattering, but so
far there is neither a defined normal form nor any particular onset of
the effect.}
\vspace*{-0.2cm}
\end{itemize}
\noindent
A great present challenge, for theory as well as for experiment, is thus
to define a normal form for strangeness production and transverse
momentum behaviour; such a reference is a prerequisite in the search
for unconventional signs.

\par

What can these results and conclusions teach us for RHIC and LHC
studies? First and foremost, I believe, they indicate that --
barring the unexpected -- higher collision energies are no
substitute for a systematic study. As we have seen, strangeness
enhancement and $p_T$-broadening occur in p-p/p-${\bar {\rm p}}$
collisions precisely at high energy. It is therefore necessary to
study p-p and p-A collisions here as well, both in experiment and
theory, before drawing any conclusions. If a given probe is not
calibrated, it cannot provide convincing results.

\par

Quite a number of different interesting experiments or experimental
components will be devoted to the search for the QGP at RHIC and LHC.
As in any scientific endeavor, we must be aware here that some results,
even if they come from an extensive and well-done experimental study,
may nevertheless simply be insensitive to deconfinement and
QGP-formation. If we insist on finding the QGP everywhere, the
scientific community at large may well doubt that we found it anywhere.

\par

Last, but not least, it seems necessary that experiments are made to
address fundamental physics questions; life is too short to refute all
wrong models. Moreover, appearently dead models quite often sprout new,
equally ugly heads, turning the battle against them into an everlasting
{\sl hydradynamics}\footnote{The Hydra was a many-headed water serpent
in Greek mythology. When one of its heads was cut off, two new heads
would appear in its place. Hercules killed it by burning the neck after
cutting off each head.}. The basic task will remain to first understand
the conventional and then identify the unconventional signs.

\bigskip

\noindent{\bf Acknowledgements}

\bigskip

The financial support by DFG contract Ka 1198/4-1, BMBF contract
06BI804/5 and GSI contract BISATT is gratefully acknowledged.

\end{document}